\documentclass[iop,apj]{emulateapj}

\usepackage{amsmath}
\usepackage{color}

\newcommand{\kl} {\,${\rm k\lambda}$~}

\begin{document}
\title{ On the Commonality of 10-30\,AU Sized Axisymmetric Dust Structures \\
 in  Protoplanetary Disks} 

\author{Ke Zhang\altaffilmark{1},   Edwin A. Bergin\altaffilmark{1}, Geoffrey A. Blake\altaffilmark{2},  L. Ilsedore  Cleeves\altaffilmark{3},  Michiel Hogerheijde\altaffilmark{4}, Vachail Salinas\altaffilmark{4}, Kamber R. Schwarz\altaffilmark{1}}

\altaffiltext{1}{Department of Astronomy, University of Michigan, 1085 S University Ave, Ann Arbor, MI 48109, USA; kezhang@umich.edu}
\altaffiltext{2}{Division of Geological \& Planetary Sciences, MC 150-21, California Institute of Technology, Pasadena, CA 91125, USA}
\altaffiltext{3}{Harvard-Smithsonian Center for Astrophysics, 60 Garden St., Cambridge, MA 02138, USA}
\altaffiltext{4}{Leiden Observatory, Leiden University, P.O. Box 9513, 2300 RA Leiden, The Netherlands}         

  \begin{abstract}
An unsolved problem in step-wise core-accretion planet formation is that rapid radial drift  in gas-rich protoplanetary disks should drive mm-/meter-sized particles inward to the central star before large bodies can form. One promising solution is to confine solids within small scale structures. Here we investigate dust structures in the (sub)mm continuum emission of four disks (TW Hya, HL Tau, HD 163296 and DM Tau), a sample of disks with the highest spatial resolution ALMA observations to date. We retrieve the surface brightness distributions using synthesized images and fitting visibilities with analytical functions. We find that the continuum emission of the four disks is $\sim$axi-symmetric but rich in 10-30\,AU-sized radial structures, possibly due to physical gaps, surface density enhancements or localized dust opacity variations within the disks. These results  suggest that small scale axi-symmetric dust structures are likely to be common, as a result of ubiquitous processes in disk evolution and planet formation. Compared with recent spatially resolved observations of CO snowlines in these same disks, all four systems show enhanced continuum emission from regions just beyond the CO condensation fronts, potentially suggesting a causal relationship between dust growth/trapping and snowlines.
 \end{abstract}


   \maketitle

\section{Introduction}
\label{sec:intro}

In the core-accretion planet formation scenario, planet formation starts with micron-sized interstellar medium grains grow into kilometer-sized planetesimals \citep{Goldreich73} --
a size interval that poses great challenges since aggregates in this size range experience significant drag from the surrounding gas and thus drift toward the central star on extremely short timescales \citep{Whipple72, Weidenschilling97}.
 One promising solution is to restrain particles in a confined area, such as a local pressure maximum in the disk \citep{Lyra08, Johansen09, Pinilla12a, Birnstie13}.  

Continuum emission at (sub)mm wavelengths provides the most direct constraints on spatial distribution of mm-sized dust grains in disks. Recent observations of some transition disks show large scale ($>$40\,AU) radial and azimuthal inhomogeneities in continuum emission, providing direct evidence of dust trapping in disks \citep[e.g.][]{Casassus13, vanderMarel13, Isella13, Perez14, Zhang14}. Such extreme inhomogeneities are commonly attributed to pressure bumps excited by giant planet(s) in the disk. However, this poses a chicken-egg dilemma on the planetesimal formation problem. Other mechanism(s) of dust trapping therefore need to be explored with higher spatial resolution observations in larger disk samples. 

The Atacama Large Millimeter/Sub-milimeter Array (ALMA) recently 
imaged the HL Tau protoplanetary disk with a superb spatial resolution of $\sim$ 5\,AU, revealing a remarkable series of dark and bright concentric rings in the continuum emission \citep{Partnership15}. The origin of these rings have been suggested to be gap opening(s) induced by embedded planets \citep{Dipierro15, Pinte16, Dong15} or changes in the dust properties near condensation fronts of dominant ices and clathrates \citep{Zhang15, Okuzumi15}. Searching for similar small scale structures in a population of disks is thus critical to study their origin and to ultimately gain an understanding of the planetesimal formation processes during gas-rich stages.

Here we  investigate the commonality of 10-30\,AU-sized dust structures in a modest sample of four protoplanetary disks with the highest spatial resolution (sub)mm continuum observations to date. 

\section{Observations and Results}
\label{sec:obs}
Our sample is composed of four protoplanetary disks:  TW Hya, DM Tau, HD 163296 and HL Tau. 
Data on TW Hya and DM Tau  were acquired out as  part of  the ALMA cycle 2 project  2013.1.00198.S, and those on HD 163296 were obtained in project 2013.1.01268.S (Salinas et al., in prep). 
Here we mainly use the public data on HL Tau \citep{Partnership15} as a test case for our data analysis methodology. A summary of observations is provided in Table 1.

\begin{deluxetable*}{cllllllll} [!htbp]
\tabletypesize{\scriptsize}
\tablecaption{Observation log and source properties \label{tab:logs}}
\tablewidth{0pt}
\tablehead{
\colhead{Source}&\colhead{$\nu_{\rm rest}$}& \colhead{$\Delta\nu$} &  \colhead{$t_{\rm int}$}        &\colhead{Beam}    & \colhead{Baseline}     &\colhead{Flux} & \colhead{Rms} &\colhead{Obs date}\\
	 		  &\colhead{(GHz)}          & \colhead{(GHz)} 	   &\colhead{(minute)} 		&\colhead{(\arcsec$\times$\arcsec (PA))} & \colhead{(k$\lambda$)}              &\colhead{(mJy)} & \colhead{(mJy beam$^{-1}$)}&\colhead{} 
}

\startdata
TW Hya	&	349	&	2.938	&	29.0	&	0.28$\times$0.28 (-10)	&	25-913	&	1415	&	0.096	&	6/15/2015	\\
	&	661	&	1.875	&	39.9	&	0.35$\times$0.20 (85)	&	33-931	&	5586	&	1.19	&	3/12/2014	\\
DM Tau	&	329	&	2.234	&	7.7	&	0.41$\times$0.33(24.9)	&	24-861	&	191	&	0.191	&	6/14/2015	\\
HD 163296	&	233	&	2	&	154.5	&	0.38$\times$0.27 (64.7)	&	19-638	&	710	&	0.017	&	7/27-29/2014	\\
HL Tau	&	233	&	8	&	280.2	&	0.035$\times$0.022(11)	&	12-11843	&	744	&	0.01	&	10/24-31/2014	\\ [0.1cm]
\hline
\hline \\[-0.25cm]
Source     &  Distance &  $M_\star$     & $L_\star$        &  $\dot{M}$    & incl     &PA &   R$_{\rm CO}$ &  Ref\\[0.1cm]
	 		  &(pc)         & ($M_\odot$/yr) &($L_\odot$) 	&($M_\odot/yr$) &(deg)    &(deg) &(AU)&\\[0.1cm]
\\[-0.3cm] \hline \\[-0.25cm]
TW Hya	&	54	&	0.55	&	0.23	&	4$\times$10$^{-10}$	&	7	&	355	&	17-23	&	1,2,3,4	\\
DM Tau	&	145	&	0.5	&	0.25	&	2$\times$10$^{-9}$	&	35	&	155	&	70$\pm$10	&	5,6	\\
HD 163296	&	122	&	2.3	&	27.2	&	7.6$\times$10$^{-8}$	&	224	&	312	&	90$^{+8}_{-6}$	&	7,8	\\
HL Tau	&	145	&	1.3	&	---	&	1$\times$10$^{-6}$	&	46.7	&	138	&	63$\pm$10	&	9			  

\enddata
\tablecomments{ References: 1.\,\citealt{Qi04}, 2.\,\citealt{Hughes11}, 3.\,\citealt{Qi13}, 4.\,Schwarz et al. submitted, 5.\,\citealt{Pietu07}, 6.\,Bergin et al. in prep, 7.\,\citealt{Rosenfeld13}, 8.\,\citealt{Qi15}, 9.\,\citealt{Partnership15} }
\end{deluxetable*}

All visibility data were calibrated in CASA (version 4.2) using scripts provided by the ALMA staff.   
The absolute uncertainty of the flux calibration is $\sim$10\%.
We performed iterative self-calibration on both the continuum emission phase and amplitude to reduce the atmospheric decoherence.

\label{sec:obs_results}

Figure~1 presents the continuum visibility profiles from the four disks as a function of the deprojected baseline length.  The baselines are generally $\sim$700\,k$\lambda$, except that those for HL Tau extend to $\sim$12,000\,k$\lambda$.  All of our sources appear to be axi-symmetric in synthesized images, as also suggested by the flat distribution of their imaginary visibility components as a function of deprojected $uv$-distance. The most important feature in the visibility profiles is that they show a wide variety of structures.  TW Hya, for example, shows a bump around 290 and 250\kl at 349 and 661\,GHz, respectively; while HD 163296 shows two bumps and a dip below zero. HL Tau has three main bumps and extensive fine structures out to the longest $uv$-distances. In contrast, DM Tau has a smooth decay and, like HD 163296, goes below zero around 400\,k$\lambda$. Since an interferometer acts as a spatial frequency filter, these visibility features can be used to reveal the detailed radial structures of the disks. 

We stress that simple disk models widely used in the literature have difficulty in reproducing the visibility features observed here (Figure 1, f-i).  A  disk with a tapered outer edge does not bring any significant visibility features. That with a sharp outer edge (even those that decay over modest radial distances) does yield features in the visibility profile, but involves a series of bumps with similar width and that occur at harmonic spatial frequencies. 
Another widely used model is a disk with a sharply truncated inner cavity. This model best fits disks that show {\em significant} negative components in their visibility profiles, and has been successful in modeling many transition disks \citep[e.g.][]{Andrews11, Zhang14, vanderMarel15}. This model, however, does not work well for the disk sample here. It yields negative components  that are too broad compared to the profiles of DM Tau and HD 163296. More specifically, the second nulls of the observed visibility profiles occur much closer to the first nulls than predicted. Furthermore, a cavity solution yields a visibility profile that monotonically decreases within the first null, while bumps are shown inside the first null in TW Hya and HL Tau. These discrepancies suggest that a sharp truncation alone cannot fit the data, and thus other structure elements are needed to explain the observations of the four disks. 

\begin{figure*}[htbp]
\includegraphics[width=7.5in]{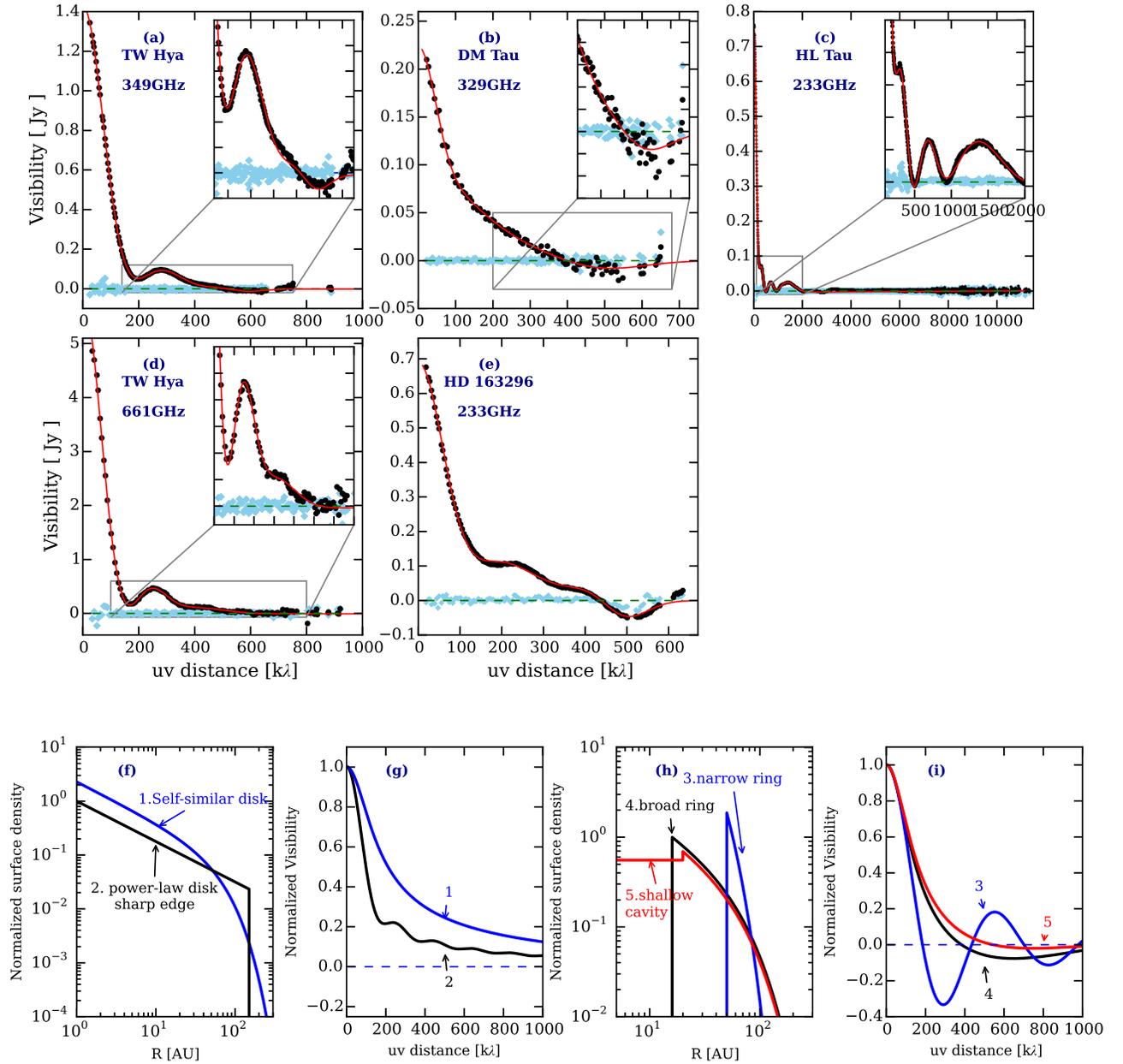}
\vspace{-0.6cm}
\caption{ Panel (a-e): Visibility profiles of continuum emission from TW Hya, HD 163296, DM Tau and HL Tau. The Real (black dots) and Imaginary (light blue diamonds) parts of the visibilities are plotted as a function of deprojected $uv$-distance, and the statistical errors are smaller than the size of symbols.  
The red lines are our best fitting models from Section 3. Panel (f,h): model surface density profiles for a disk at 140\,pc; Panel (g, i):  visibility profiles of the simple disks models in (f,h).  The data of panel (a-e) are available for download from arXiv source. }
\label{fig:f_r}
\vspace{-0.15cm}
\end{figure*}

\section{Modeling}

As discussed above, the simplest physical models fail to characterize the observed visibility profiles. 
More importantly, there is no coherent physical model available for the surface density profile of mm-sized particles, because it may differ significantly from that of a viscously evolving gas disk due to radial drift and various pressure trapping mechanisms \citep{Chiang10, Andrews15}. Due to these uncertainties, we employ an empirical approach to characterize spatial structures in the surface brightness distributions and discuss the possible origin of the observed disk structures in Section 4.

\subsection{Deconvolution in interferometric observations} 

The retrieval of  source intensity distributions from visibilities is essentially a deconvolution process, due to discrete sampling on the $uv$-plane. Critically,
without prior information, the deconvolution has no unique solution because the fine structures in the source intensity distribution correspond to unsampled high spatial frequency components that can have a wide range of amplitudes \citep{Cornwell99}.

The most widely used deconvolution method in heterodyne interferometry is the CLEAN algorithm \citep{Hogbom74, Clark80},
in which the final deconvolved image is a summation of a number of point sources convolved with a CLEAN beam (usually a Gaussian). This approach suppresses the highest spatial frequencies in the data and results in a smeared image. 

Another common way to derive source intensity is the so-called \textit{modeling fitting} approach \citep{Pearson99}. Here, the observed visibilities are reproduced with a {\em parametric} model of the source intensity distribution. Advantages of this method include the utilization of the full spatial frequency information in data and straightforward error estimation. A significant drawback is that the possible form of models that fit the data may not be unique. Thus the choice of a model function requires physical justification.

Here we retrieve the radial surface brightness distributions of our disk sample using both the image (CLEAN) and model fitting approaches. 

\subsection{Model fitting approach}
For circularly symmetric disk emission, the link between the deprojected $uv$-distance and radial brightness distribution is a Hankel transform \citep{Pearson99}: 

\begin{eqnarray}
\label{eq:hankel}
u' &=& (u {\rm cos}\phi-v {\rm sin}\phi)\times {\rm cos}\,i  \\
v'& =&  u {\rm sin}\phi+v {\rm cos}\phi\\
V (\rho) &=& 2\pi\int_0^\infty I_\nu (\theta) \theta J_0(2\pi\rho \theta) d\theta,
\end{eqnarray}

\vspace{-0.25cm}

\noindent where $i$ and $\phi$ are the disk inclination and position angle, $\rho$=$\sqrt{u'^2+v'^2}$ is the deprojected $uv$-distance in units of $\lambda$, $\theta$ is the radial angular scale from the disk center, and $J_0$ is a zeroth-order Bessel function of the first kind. 

Here we adopt an analytical function for $I(\theta)$ that is inspired by the multi-peak features seen in the visibility profiles (Figure~1). A peak in visibility indicates that some spatial frequencies, corresponding to some particular spatial scales, have more contribution than other scales. Specifically, we model the disk surface intensity distribution $I(\theta)$ with a group of Gaussian functions, each of which is modulated by a sinusoidal function with a spatial frequency of $\rho_i$ (eq.~\ref{eq:i_theta}).  The number of Gaussians is decided by the number of distinctive peaks in the visibility profile, and $\{ a_0, \sigma_0, a_i, \sigma_i, \rho_i$\} are free parameters. 
Thus, we choose

\begin{multline}
\label{eq:i_theta}
I(\theta) = \frac{a_0}{\sqrt{2\pi}\sigma_0} exp \left(-\frac{\theta^2}{2\sigma_0^2}\right) \\
+\sum_{i} {\rm cos} (2\pi\theta \rho_i)\times \frac{a_i}{\sqrt{2\pi}\sigma_i} exp \left(-\frac{\theta^2}{2\sigma_i^2}\right)
\end{multline}

This analytic function is empirical but consistent with realistic disk emission in several aspects.  It insures that $I(\theta)$ goes to zero at infinity. Further, the amplitudes of components associated with unsampled high spatial frequency go to zero quickly, meaning no fine structure information is added.  Because real disk emission never becomes negative, the amplitude of V$(\rho)$ should gradually decay with spatial frequency.   Thus, adding new visibility data with higher spatial frequency coverage will not change the known structures in $I(\theta)$ drastically.  This approach is suitable for $I(\theta)$ functions without hard edges and its utility is supported by the fact that no harmonic features associated with sharply truncated disks are found in the four disks. As the largest recoverable spatial scale (determined by the shortest baseline) is significantly larger than the disk emission, the total flux recovered from the analytic function is conserved. 

After initial fitting, we find that HD 163296 and DM Tau show a flux decrement inside $\sim$ 20\,AU.  To investigate if including a sharp inner edge would change the derived disk structures, we run additional models for the two disks, by adding two extra free parameters to simulate an inner cavity --- a sharp inner edge R$_{\rm cav}$ and a depletion factor $\delta$ ($0\le\delta\le1$). We assume the source intensity is flat inside the cavity since fine structures inside 20\,AU are unresolved.

We use the Levenberg-Marquardt $\chi^2$ minimization algorithm to search for the optimal value of free parameters. The initial guesses of the \{$\rho_i$\} are the centers of peaks in visibility profiles and values of \{$a_i$\} can be either positive or negative. The integral in eq.~(\ref{eq:i_theta}) is solved using a step size of 0.1\,AU from 400\,AU. 

\begin{figure*}[htbp]
\vspace{-0.4cm}
 \includegraphics[width=7.5in]{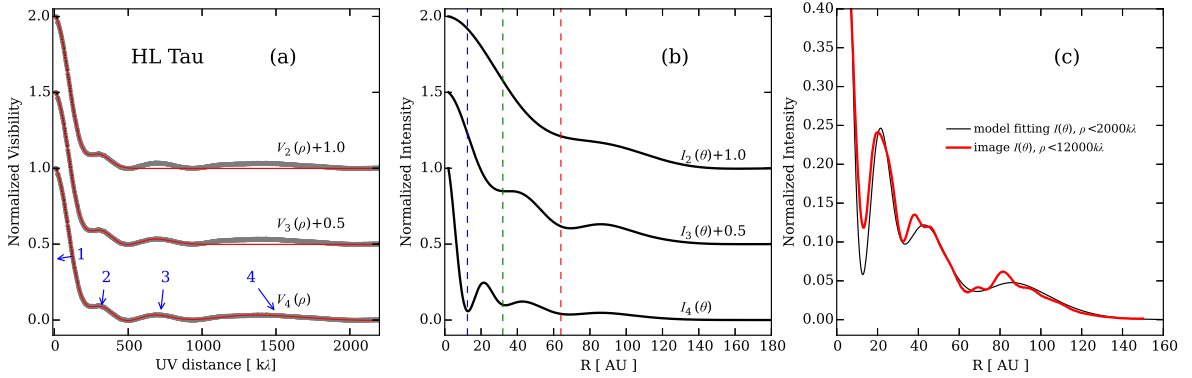}

\caption{HL Tau as an example of  the model fitting approach. Panels (a) and (b) show how the derived surface intensity profile $I(\theta)$ changes when visibilities from larger $uv$-distances are included. The functions $V_2$, $V_3$ and $V_4$ in (a) correspond to the number of Gaussian functions in the model fitting, and the three vertical dashed lines in (b) highlight the centers of three major disk gaps reported by \citealt{Partnership15}. (c) A comparison of the surface brightness derived from fitting the $\rho<$2000\kl visibilities and that resulting from a CLEANed image using all $\rho<$12000\kl data. }
\label{fig:example}
\vspace{-0.15cm}
\end{figure*}

 Using HL Tau as an example, Figure~\ref{fig:example}\,(a-b) show how including more visibilities changes the derived radial intensity distribution.  In particular, by fitting only two Gaussians out to 500\,k$\lambda$, we find a broad gap around 60\,AU. Extending the data to 1000\,k$\lambda$ then demands fits with three Gaussians, and a second but narrower gap is found around 30\,AU. When we include data within 2000\,k$\lambda$, an innermost gap is seen at $\sim$13\,AU.  

\subsection{Image approach}

Here we obtain radial intensity profiles from the CLEANed images directly. 
The visibilities are deprojected using eqs. (1-2), and deconvolved through the CLEAN algorithm using uniform weighting, and restored with a synthesized Gaussian beam. We then derive an azimuthally averaged $I(\theta)$ from the images. An illustrative comparison of radial intensity profiles of HL Tau is plotted in Figure~2(c).  Clearly, the three major gaps (13, 32 and 63\,AU) are reproduced nicely by the model fitting approach using the $\rho\le 2000$\,k$\lambda$ data, as compared with the $I(\theta)$ derived from a CLEANed image based on visibilities of $\rho_{\rm max}\sim12000$\,k$\lambda$.

\subsection{Results}

The retrieved radial intensity distributions of the four disks from both the image and modeling fitting approaches are presented in Figure~3. Consistent results are found over larger scales, but (as expected by the HL Tau example) the modeling fitting results clearly show more detailed structures. The continuum emission from all four disks is rich in radial structures with a typical length scale of 10-30\,AU.  For TW Hya, both approaches show that the 349 and 661\,GHz emission has a turning point in the slope around 25\,AU, followed by a plateau and then a gradual decay out to $\sim$70\,AU.  \citet{Nomura15} recently reported similar structures in the 336\,GHz continuum emission of TW Hya.  For HL Tau, its known major gaps at 13, 32 and 63\,AU are well recovered. HD 163296 and DM Tau are two disks that show signs of central flux decrement in our modeling (see also DM Tau in Andrews et al. 2011). For these two, we show the best fitting results of two analytical models: (1) a smooth disk (no sharp inner edge), and (2) a disk with a sharp inner edge (R$_{\rm cav}$, $\delta$). Both types of models retrieve consistent structures beyond $\sim$ 30\,AU -- HD 163296 has two depressed zones centered near 55 and 100\,AU, and  DM Tau has a shallowly depressed emission zone around 70\,AU. The two depressed zones in HD 163296 are also noticeable in its synthesized image along the beam minor axis. The two types of models give slightly different structures in the central region of HD 163296 and DM Tau, but both suggest that the central regions are probably just shallowly depleted. It is likely that there are unresolved emission from the central regions and the flux decrement is possibly due to gaps rather a central cavity.  

It is important to note that our proposed solutions are consistent with the data but other disk structures may also be possible. This is a nature of deconvolution that the solution is not unique without a restrictive physical framework, a framework we currently lack in interpolating mm-sized particle distribution in disks. Under this condition, the simplest solutions are preferred.  Our proposed solutions belong to the the simplest group since they are the smoothest models that fit the data (least high spatial frequency components needed).

\begin{figure*}[t]

\hspace*{-0.25in}
 \includegraphics[width=8.0in]{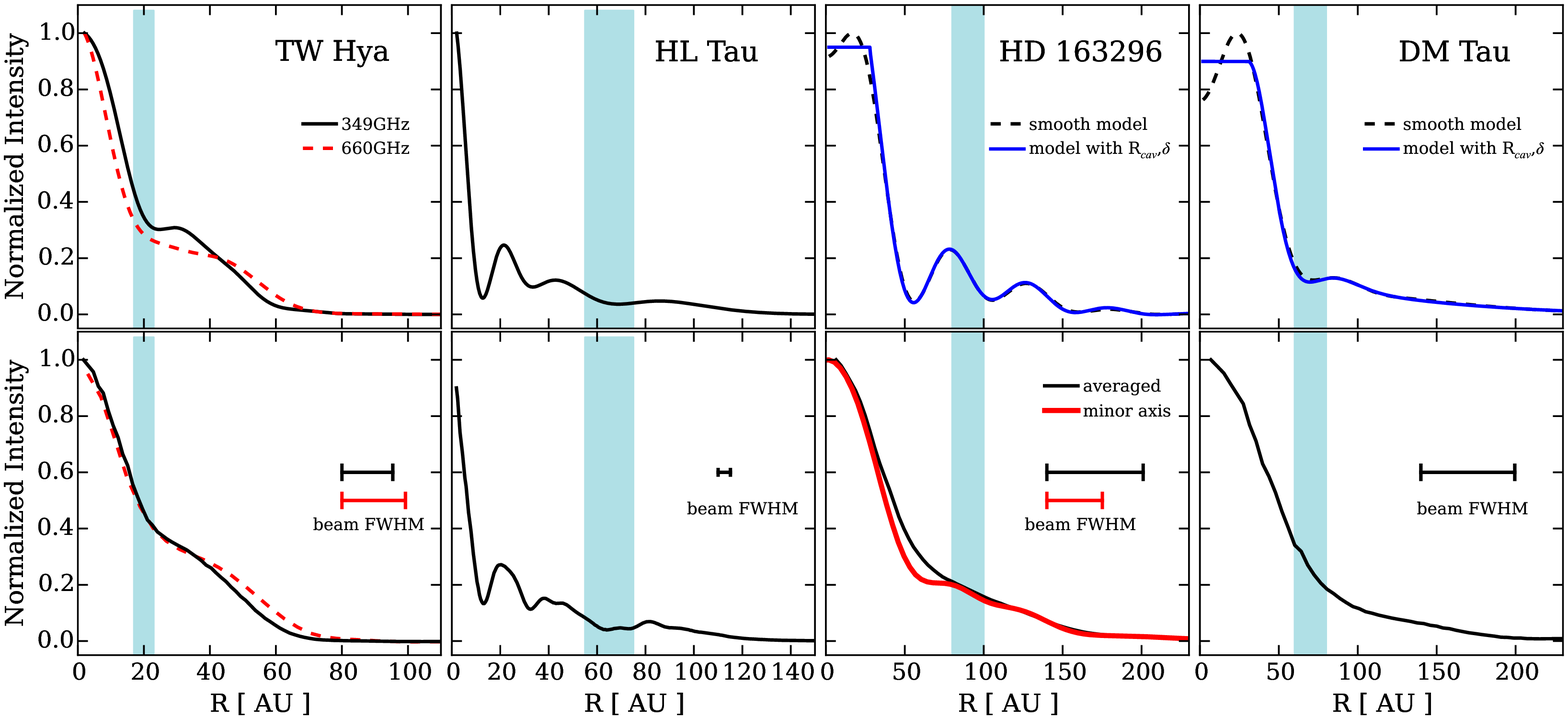}

\caption{Radial surface brightness profiles of the four disks, derived from modeling fitting (upper row) and synthesized images (bottom row). The results of each disk are derived from the same visibilities, except for HL Tau, for which the modeling fitting result uses visibilities with a maximum $uv$ distance that is much shorter than those used in the image ($\rho_{\rm max}\sim2000$\,k$\lambda$ v.s. 11843\,k$\lambda$). For the modeling fitting results of HD 163296 and DM Tau, we show two $I(\theta)$ profiles derived from a smooth model as well as a model with a sharply truncated central depletion zone.  The light blue regions highlight the expected mid-plane CO snowline region, including uncertainties.}
\label{fig:f_r}

\end{figure*}

\begin{figure*}[!htbp]
\vspace{0.4cm}
\hspace*{-0.25in}
 \includegraphics[width=8in]{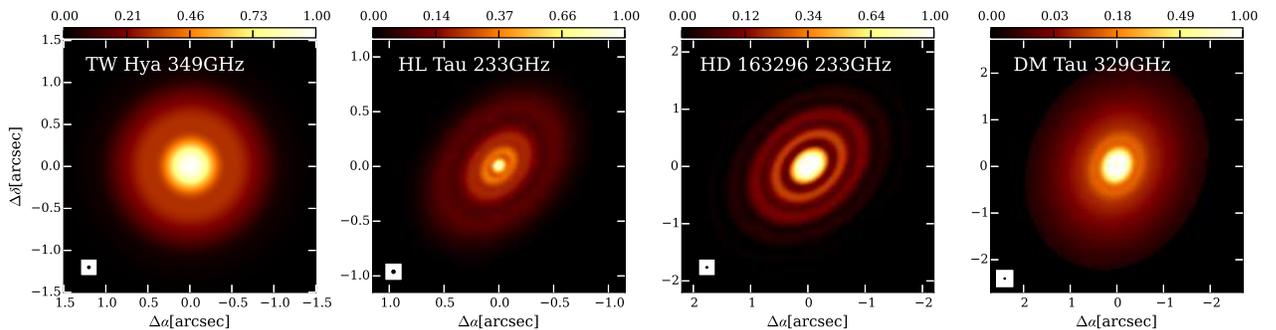}
\caption{Prediction of long-baseline ALMA images (at 0.\arcsec035 resolution). The surface brightness distributions are based on our best model fitting of observed visibilities (Figure 3, top row). The intensity is normalized to the brightest pixel in each image. }
\label{fig:example}
\vspace{0.15cm}
\end{figure*}

\section{Discussion}

\subsection{Origin of disk structures}

The richness of small scale features in the surface brightness profiles shown in Figure~\ref{fig:f_r} is striking. 
One possibility is that they are results of pressure bumps in protoplanetary disks. Possible mechanisms proposed for generating pressure bumps include: zonal flows \citep{Johansen09b, Simon12}, planet-disk interaction \citep{Lyra09}, or a (water) snowline-induced jump in surface density/ionization degree \citep{Kretke07}.

On the other hand, disk structures in (sub)mm continuum emission may also be produced by spatial variations in the dust opacity $\kappa_\nu(R)$.  For example, grain growth itself, will alter the dust opacity, and if localized, the change in opacity can mirror the effects of a change in the surface density profile.    As but one example of a local effect that dust growth near the water snowline from millimeter to decimeter-sized pebbles is possible on a timescale of only 1000 years \citep{Ros13}.  

\subsection{Commonality of smooth disk structures}
So far, ALMA high spatial resolution observations have been only carried out for several well-studied classical T Tauri or Herbig star disks or transition systems (disks with a large central dust cavity). The majority of transition disks observed show some axi-asymmetry and sharp edges \citep{vanderMarel15}.  The fraction of transition disks is estimated to be $\sim$10-20\% in nearby star-forming regions, based on the spectral energy distribution statistics  \citep{Kim09, Merin10}; a greater fraction of $\sim$ 30\% \citep{Andrews11} is given based on resolved sub-mm imaging. Nevertheless, current statistics indicates that disks with a large cavity ($>$20\,AU) are probably not the majority of disks.

The four disks studied here perhaps provide a better match to the structure of the majority of disks, i.e.,  `full' disks or disks with small holes. One prediction from our sample is that circularly symmetric and smoothly varying structures with $\sim$ 10-30\,AU scale lengths are likely to be more common in disks in which the central cavity is either small or absent, as compared to structures such as the sharply truncated narrow rings or edges often seen in disks with a large central cavity.
\subsection{Correlation between CO snowlines and enhanced continuum emission}

With sufficient spatial resolution to resolve CO snowlines, current ALMA observations have enabled the first direct investigations of the role of condensation fronts in the evolution of solids in nearby protoplanetary disks. 

So far, N$_2$H$^+$ cation and C$^{18}$O emission have been used as two independent tracers of the mid-plane CO snowline.  The two tracers show consistent results in TW Hya, where the mid-plane CO snowline is found to lie at 17--23~AU (\citealt{Qi13}, Schwarz et al. submitted), and for HD 163296 where R$_{\rm CO}$=90$^{+8}_{-6}$\,AU (\citealt{Qi15}). 
Since no similar observations are available for HL Tau and DM Tau, here we adopt CO snowline radii based on radiative transfer models of the two disks, which yield R$_{\rm CO clatherate}$ = 63$\pm$10\,AU for HL Tau \citep{Menshchikov99, Zhang15} and R$_{\rm CO }$ = 70$\pm$10\,AU for DM Tau (Bergin, in prep).

Figure~3 presents the mid-plane CO snowline locations on top of the surface brightness distributions in the four disks. The data suggest that there is possibly a relationship between the snowline location and enhancements in the continuum emission.
 
\subsection{Efficient selection of disk candidates for long baseline ALMA observations }

The HL Tau observations demonstrate that ALMA long baseline observations are critical to revealing structures within the nominal planet-forming disk radii ($<$30\,AU). However,  a preliminary selection of sources is desired due to the long integration times and excellent atmospheric stability needed for long baseline observations. The modeling fitting analysis above shows that distinctive features at visibility profiles are useful predictors of fine structures in disks.  To search for 10\,AU scale structures in disks of nearby star forming regions ($\sim$140\,pc),  we propose that an initial survey with $\sim$ 2000\kl baseline (e.g. 2.6\,km at 1.3\,mm) should be sufficient to select sources with significant features in deprojected visibility profiles to then be imaged with full baseline (16\,km) ALMA observations. In Figure\,4, we show simulated long-baseline ALMA images  of our sample. The significant contrast in the HD 163296 image is induced by the visibility null near 450\,k$\lambda$. The structures proposed here can easily be confirmed or ruled out in long baseline observations, and similarly the potential association between CO snowlines and continuum emission enhancements.

\vspace{0.2in}
This work was supported by funding from the National Science Foundation, grants AST-1344133 (INSPIRE) and AST-1514670.
The authors thank John Monnier for discussions on model fitting approaches to visibility data. The National Radio Astronomy Observatory is a facility of the National Science
 Foundation operated under cooperative agreement by Associated Universities, Inc.
This paper makes use of the following ALMA data sets: ADS/JAO.ALMA\#2011.0.00015.SV, ADS/JAO.ALMA\#2013.1.00198.S \& ADS/JAO.ALMA \#2013.1.01268.S. ALMA is a partnership of ESO (representing its member states), NSF (USA) and NINS (Japan), together with NRC (Canada), NSC and ASIAA (Taiwan), and KASI (Republic of Korea), in cooperation with the Republic of Chile. The Joint ALMA Observatory is operated by ESO, AUI/NRAO and NAOJ.


\end{document}